\documentclass{siamltex}

\usepackage[breaklinks,colorlinks=true,linkcolor=blue,citecolor=red]{hyperref}

\usepackage{amsfonts} 
\usepackage{amssymb,amsmath} 
\usepackage{graphicx,color}
  
\newcommand{\eps}{\varepsilon} 
 
\newcommand{\RR}{{\mathbb R}} 
\newcommand{\EE}{{\mathbb E}}

\DeclareMathAlphabet{\itbf}{OML}{cmm}{b}{it}

\newcommand\bx{{\itbf x}}

\newcommand\by{{\itbf y}}
\newcommand\bz{{\itbf z}}

\newcommand{\qed}{\hfill $\Box$ \medskip}

\def\qed{\hfill {\small $\Box$} \\}

\begin{document}  
\title{Passive communication with ambient noise} 

\author{Josselin Garnier\footnotemark[1]}   

\maketitle

\renewcommand{\thefootnote}{\fnsymbol{footnote}}

\footnotetext[1]{CMAP, CNRS, Ecole polytechnique, Institut Polytechnique de Paris, 91128 Palaiseau Cedex, France, and INRIA, France
{\tt josselin.garnier@polytechnique.edu}}

\renewcommand{\thefootnote}{\arabic{footnote}}

\maketitle

\begin{abstract}
Motivated by applications to wireless communications,
this paper addresses the propagation of waves transmitted by ambient noise sources and interacting with 
metamaterials.
We discuss a generalized Helmholtz-Kirchhoff identity that is valid in dispersive media 
and we characterize the statistical properties of the empirical cross spectral density of the wave field.
We can then introduce and analyze an original communication scheme between two passive arrays 
that uses only ambient noise illumination. The passive transmitter array does not transmit anything but it is a tunable metamaterial surface that can modulate its dispersive properties and encode a message in the modulation.
The passive receiver array made of two receivers that are half-a-wavelength apart from each other can decode the message from the empirical cross spectral density of the wave field.
\end{abstract}



\section{Introduction}

\begin{figure}
\begin{center}
\vglue-0.8cm
\begin{tabular}{cc}
\hspace*{-1.6cm}
\includegraphics[width=8.5cm]{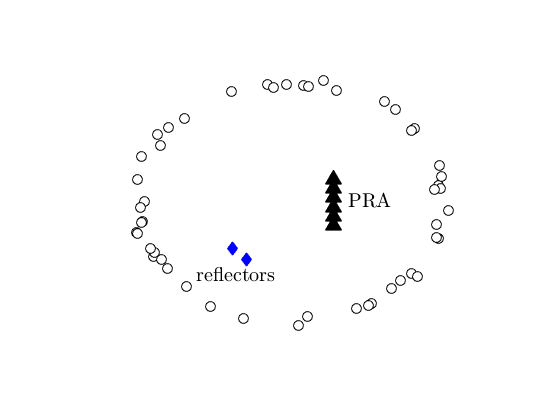}
\hspace*{-2.cm}
&
\includegraphics[width=8.5cm]{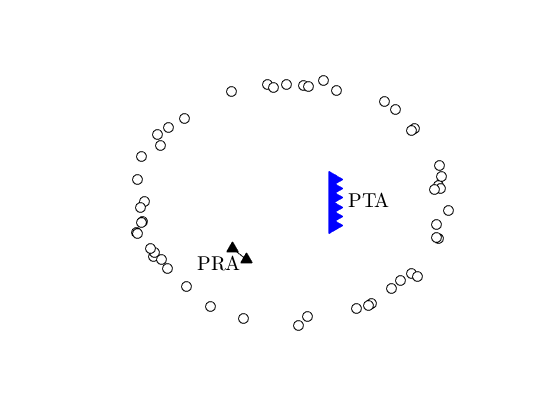}
\hspace*{-1.6cm}
\end{tabular}
\end{center}
\vglue-1.2cm
\caption{Left picture: passive imaging setup.
The circles are the distant opportunistic sources.
The large PRA has many receivers.
The aim is to localize the reflectors embedded in the medium from the signals recorded by the PRA.
Right picture: passive communication setup.
The circles are the distant opportunistic sources.
The large PTA has many elements with tunable dispersive properties.
The aim is to transmit information from the large PTA to the small PRA 
made of a limited number of receivers.
}
\label{fig:setup0}
\end{figure}

The use of ambient noise illumination for passive imaging (i.e. imaging methods using only passive receiver arrays)
has been extensively studied in the last fifteen years \cite{noisebook,helin18,wap10}, with successful applications to seismology for instance \cite{agaltsov2,schuster,shapiro}.
In this paper we propose another use of ambient noise illumination. 
In analogy with passive imaging, the aim is to perform a task - here, communication -
using only passive arrays and ambient noise illumination or opportunistic sources.
As in passive imaging, a fundamental formula that derives from Green's identities drives the design and the analysis of the proposed communication method.
Surprisingly, our analysis reveals that the term of this formula that is useful for communication is the one that is not useful for imaging, and vice-versa.
Based on this analysis, we propose and study a way to communicate from a passive transmitter array (PTA) to a passive receiver array (PRA)
by using ambient noise illumination only.
The PTA is a metamaterial surface that consists of devices that can be tuned to change their dispersive properties and to impose phase shifts on the scattered waves \cite{cui14,kaina}.
The PRA consists of two (or more) standard receiver antennas that record the wave field and compute the cross spectral density between the recorded signals
(see Figure \ref{fig:setup0}).

The theoretical motivation of this paper comes from the Helmholtz-Kirchhoff identity and its variants.
The standard Helmholtz-Kirchhoff identity states that the time-harmonic Green's function $\hat{G}$ of the Helmholtz equation
in an inhomogeneous medium satisfies
\begin{align}
\lim_{L \to +\infty} 
\frac{\omega}{c_o}   \int_{\partial B({\bf 0},L)} 
\overline{ \hat{G}( \omega,\bx,\bz)}  \hat{G}( \omega ,\by,\bz)  d \sigma(\bz) 
=  {\rm Im} \big( \hat{G}(\omega, \bx,\by) \big),
\label{eq:standardhk}
\end{align}
where $\partial B({\bf 0},L)$ is the surface of the ball with center at ${\bf 0}$ and radius $L$ and $c_o$ is the background velocity at infinity.
This identity has important consequences as it can be used to explain the refocusing properties
of time-reversal experiments \cite{noisebook} and to determine the resolution properties of active imaging methods \cite{ammarigarnier}.
It also explains why passive ambient noise imaging can work, because it shows that the cross covariance function of the signals transmitted by distant noise sources and recorded by two receivers at $\bx$ and $\by$ (which involves the left-hand side of (\ref{eq:standardhk})) is related to the Green's function between these two receivers (which is in the right-hand side of (\ref{eq:standardhk})) \cite{noisebook}.
The formula (\ref{eq:standardhk}), however, holds true only when the medium is non-dispersive or when it is dispersive but invariant by time reversal.
In the presence of dispersive inclusions, such as metamaterials or resonant particles,
the formula contains additional terms (see Appendix \ref{sec:app}) that have very interesting behaviors and open the way to new studies and new applications.
Indeed, these terms depend on the imaginary parts of the frequency-dependent coefficients of the medium.
As a consequence, we will see that, by controlling some small resonant devices,  
it is possible to generate small but special scattered waves from an arbitrary illumination, and these special wave components 
can be detected by a pair of receivers, without the necessity for the receivers to know the illumination or the positions of the resonant devices.

The application we have in mind is wireless communications.
Recently, smart reflecting surfaces or tunable metasurfaces have been proposed as tools for controlling wave propagation and achieving strong focusing on receivers in the microwave domain \cite{cui14,kaina} and in the acoustic domain \cite{ma18}. 
A tunable metasurface is composed of a large number of passive elements that can be tuned to reflect the incident wave with an adjustable phase shift. By tuning the phase shifts of the elements of the metasurface adaptively, the reflected waves can be made to interfer constructively at a nearby receiver for instance, which makes it possible to enhance the signal power on the receiver and to improve the communication performance \cite{wu20}.
Other applications have been proposed to localization of reflectors  \cite{delhougne} or covert communication \cite{lu}.
The main point is that the tunable metasurface is passive, it does not involve any active transmission. It is, therefore,
very attractive for the next 5G or 6G generation and the Internet of Things (IoT) \cite{basar19}.
Taking inspiration from backscatter communication systems \cite{liu13,xu18}, 
in which an antenna modulates its impedance to encode information in an incoming signal  emitted  by  a  nearby transmitter,
the possibility to use opportunistic sources and tunable metasurfaces is explored for wireless communications in \cite{zhao20}.
This requires 1) to deploy (at least) two coherent receivers (the master receiver and the slave receiver) to acquire the wireless signals, 2)  to localize the two receivers relatively to the metasurface, and 3) to focus the transmitted signal on the master receiver and to make it invisible from the slave receiver.
In this paper, we show that, using ambient noise illumination, a tunable metasurface can be used as a PTA to transmit information to a PRA in a fully passive way, without the need to localize the receiver or to focus the signal. 
We clarify the conditions under which such a communication scheme can be implemented and we show that it is robust with respect to measurement noise and clutter noise.

The paper is organized as follows.
In Section \ref{sec:ambient} we introduce the empirical cross spectral density
that a pair of receivers can compute from the signals transmitted by distant ambient noise sources through an arbitrary dispersive medium.
In Section \ref{sec:green} we describe the structure of the Green's function 
in a medium with resonant inclusions or particles.
In Section \ref{sec:passivecomm} we describe the passive communication scheme and study its properties in terms of transmission rate and robustness.

\section{Waves transmitted by ambient noise sources}
\label{sec:ambient}
We consider the solution $u(t,\bx)$ of the wave equation in a three-dimensional 
inhomogeneous and dispersive medium.
The inhomogeneities are assumed to be localized in a bounded region of $\RR^3$.
The Fourier transform $\hat{u}(\omega,\bx)$ of the wave field $u(t,\bx)$
$$
\hat{u}(\omega,\bx) = \int_\RR u(t,\bx) e^{i \omega t} dt
$$
 satisfies the Helmholtz equation
with the Sommerfeld radiation conditions  (see Appendix~\ref{sec:app}).
It has the integral representation:
$$
\hat{u}(\omega,\bx) = \int_{\RR^3}\hat{G}(\omega,\bx,\by)\hat{s}(\omega,\by) d\by ,
$$
where $\hat{G}(\omega,\bx,\by)$ is the time-harmonic Green's function,
that is, the fundamental solution of the Helmholtz equation with 
a point source at $\by$ (see Appendix \ref{sec:app}).
The term $\hat{s} (\omega,\bx)$ models a random field of noise sources.
It is the Fourier transform of a zero-mean stationary (in time) random process
with cross covariance function
\begin{equation}
\label{def:autocorn}
\langle s(t_1,\by_1) s(t_2,\by_2) \rangle =  F (t_2-t_1)  \Gamma(\by_1,\by_2) .
\end{equation}
Here $\langle \cdot \rangle$ stands for statistical average with respect to
the distribution of the noise sources.
For simplicity we consider that the process $s(t,\bx)$ has Gaussian statistics.

The time distribution of the noise sources is characterized 
by the correlation function $F(t_2-t_1)$, 
which is a function of $t_2-t_1$ only because of time stationarity.
The function $F$ is normalized so that $F(0)=1$.
The Fourier transform $\hat{F}(\omega)$ of the time correlation function $F(t)$
is a nonnegative, even, real-valued function proportional to the power spectral
density of the sources by Wiener-Khintchine theorem.

\subsection{The cross covariance function of the wave field}

The spatial distribution of the noise sources is characterized 
by the covariance function $\Gamma(\by_1,\by_2)$.
We assume that the  random process $s$ is delta-correlated in space:
\begin{equation}
\label{relieRtheta}
 \Gamma(\by_1,\by_2) = {K} ( \by_1) \delta(  \by_1 -\by_2 ) .
\end{equation}
The function ${K}$ then characterizes the spatial support of the sources.
Extensions to spatially correlated sources are possible as seen in \cite{reso}.

\begin{proposition}
\label{prop:1}
$u(t,\bx)$ is a stationary in time Gaussian process with mean zero and cross covariance function
\begin{equation}
\label{realfourier}
C^{(1)} (\tau,\bx_r,\bx_{r'}) :=\left< u(t,\bx_r)u(t+\tau,\bx_{r'}) \right>
\end{equation}
given by
\begin{align}
\label{eq:expressC1}
C^{(1)} (\tau,\bx_r,\bx_{r'}) = \frac{1}{2\pi}
\int_{\RR} \hat{F}(\omega) \hat{Q}(\omega,\bx_r,\bx_{r'})  e^{-i\omega \tau} d\omega  , \\
\hat{Q}(\omega,\bx_r,\bx_{r'}) = \int_{\RR^3}
   {K}(\by )
 \overline{\hat{G}(\omega,\bx_r,\by )} \hat{G}(\omega,\bx_{r'},\by)  d\by .
 \label{def:hatQ}
\end{align}
\end{proposition}

\noindent
{\it Proof.}
The wave field $u(t,\bx)$ has Gaussian statistics and mean zero since it is a linear transform of the source $s (t,\bx)$ that has Gaussian statistics.
We can express the cross covariance function $C^{(1)}$ in the frequency domain:
\begin{align*}
C^{(1)} (\tau,\bx_r,\bx_{r'}) = \frac{1}{(2\pi)^2}
\iint_{\RR^3 \times \RR^3  \times \RR \times \RR}
 \overline{\hat{G}(\omega,\bx_r,\by )} \hat{G}(\omega',\bx_{r'},\by')  e^{i\omega t -i \omega'(t+\tau)} \\
\times \left< \overline{ \hat{s}(\omega,\by)}\hat{s}(\omega',\by') \right>
 d\by d \by' d\omega  d\omega'
 ,
\end{align*}
and use the form of the cross covariance function of the noise sources:
\begin{equation}
\label{eq:mom2s}
\left< \overline{ \hat{s}(\omega,\by)}\hat{s}(\omega',\by') \right>
=2\pi \hat{F}(\omega) \delta(\omega-\omega')  K(\by) \delta(\by-\by'),
\end{equation}
to get the desired result.
\qed

Proposition \ref{prop:1} gives the full statistical characterization of the wave field.
In practice, of course, we do not measure the cross covariance function
$C^{(1)}$ that is a statistical average.
In ambient noise imaging the relevant quantity that is measured is the empirical cross covariance function \cite{garpapa09,noisebook}:
$$
C_T(\tau,\bx_r,\bx_{r'}) = \frac{1}{T} \int_0^T u(t,\bx_r)u(t+\tau,\bx_{r'}) dt .
$$
It is appropriate for imaging purposes and for broadband noise illumination.
It is a statistically stable quantity when $T$ is large enough, i.e., its standard deviation is much smaller than its expectation. As a result, a typical value of $C_T$ is close to the expectation, which is equal to the cross covariance function $C^{(1)}$.
In the applications we have in mind, where the bandwidth is rather narrow and
the goal is to transmit information, 
another statistically stable quantity turns out to be more appropriate
and it is introduced in the next section.

\subsection{The empirical cross spectral density}
We have seen that the field $u(t,\bx)$ is a stationary in time Gaussian process with mean zero. Its distribution can be characterized by its cross covariance function $C^{(1)} (\tau,\bx_r,\bx_{r'}) $.
Equivalently, its distribution can be characterized by its cross spectral density, which is the Fourier transform of the cross covariance function:
\begin{align}
\nonumber
S^{(1)}(\omega,\bx_r,\bx_{r'}) &= \int_\RR C^{(1)}(\tau,\bx_r,\bx_{r'}) e^{i \omega \tau} d\tau
\\
&=
\int_\RR
\left< u\big(t-\frac{\tau}{2},\bx_r\big)u\big(t+\frac{\tau}{2},\bx_{r'}\big) \right>
e^{i \omega \tau} d\tau.
\end{align}
In practice, of course, we do not measure the cross spectral density $S^{(1)}$ that is a statistical average.

Let $T,T'>0$. We introduce the empirical cross spectral density (ECSD)
of the signals recorded at $\bx_r $ and $\bx_{r'}$:
\begin{equation}
\label{def:ST}
S_{T}(\omega) = 
   \iint_{\RR^2} u\big( t -\frac{\tau}{2}, \bx_r\big)u\big( t +\frac{\tau}{2}, \bx_{r'}\big)
e^{ i\omega \tau} \phi_T(t) \psi_{T'}(\tau) dt d\tau  ,
\end{equation} 
where we have used the cut-off functions
\begin{equation}
\phi_T(t)= \frac{1}{T} \phi\Big(\frac{t}{T}\Big), \quad \quad
\psi_{T'}(\tau)= \frac{1}{T'} \psi\Big(\frac{t}{T'}\Big), 
\end{equation}
with $\psi$ and $\phi$ two nonnegative-valued, even, integrable functions with
nonnegative-valued Fourier transforms satisfying
 $\int_\RR \phi(t) dt=\int_\RR \psi(t) dt=1$.
We may think at $\phi(t), \, \psi(t)= \pi^{-1/2} \exp( -t^2)$ or $(1-|t|)_+$ for instance.

\begin{proposition}
\label{prop:momECSD}
The mean of the ECSD is independent of $T$ and equal to
\begin{align}
\nonumber
\left< S_{T}(\omega) \right> &
= \frac{1}{2\pi} 
\int_\RR
S^{(1)} (\omega_1,\bx_r,\bx_{r'}) 
\hat{\psi}\big( T'(\omega-\omega_1)\big)
d\omega_1 
\\
&= 
\frac{1}{2\pi} 
\int_\RR
\hat{Q}(\omega_1,\bx_r,\bx_{r'}) \hat{F}(\omega_1) 
\hat{\psi}\big( T'(\omega-\omega_1)\big)
d\omega_1 ,
\label{eq:meanST1}
\end{align}
where the kernel $\hat{Q}$ is defined by (\ref{def:hatQ})
and $\hat{\psi}$ is the Fourier transform of $\psi$.
\\
The variance of the ECSD 
\begin{align}
{\rm Var}(S_T(\omega)) = \left< \big| S_T(\omega) - \left< S_{T}(\omega) \right> \big|^2 \right>
\end{align}
is equal to
\begin{align}
\nonumber
& {\rm Var}(S_T(\omega)) =
 \frac{1}{4\pi^2} 
 \iint_{\RR^2}
\Big[ 
\hat{Q}(\omega_1,\bx_r,\bx_r) \hat{Q}(\omega_2,\bx_{r'},\bx_{r'}) 
\Big|\hat{\psi} \Big(   {T'} \big( \omega- \frac{\omega_1+\omega_2}{2} \big)\Big) \Big|^2
\\
\nonumber
&\quad + \hat{Q}(\omega_1,\bx_r,\bx_{r'}) \hat{Q}(\omega_2,\bx_{r'},\bx_r) 
\hat{\psi} \Big(   {T'} \big( \omega- \frac{\omega_1+\omega_2}{2} \big)\Big) 
\overline{\hat{\psi}}\Big(   {T'} \big( \omega + \frac{\omega_1+\omega_2}{2} \big)\Big) 
\Big] 
\\
&\quad \times 
\big|\hat{\phi} \big(   {T} ( \omega_1-\omega_2)\big) \big|^2 \hat{F}(\omega_1)
\hat{F}(\omega_2)
 d\omega_1 d\omega_2 .
\label{eq:varST1}
\end{align}
\end{proposition}
{\it Proof.}
The expression (\ref{eq:meanST1}) of the mean of the ECSD is obtained from (\ref{eq:expressC1}).
We express the second-order moment of  the ECSD $S_T(\omega)$ as a multiple integral that involves the product of four Green's functions
and the fourth-order moment of the source term $\hat{s}$.
By using Eq.~(\ref{eq:mom2s}) and the Isserlis formula satisfied by the Gaussian process $\hat{s}$:
\begin{align*}
 &\langle \overline{\hat{s}(\omega_1,\by_1)} \hat{s}(\omega_2,\by_2) \hat{s}(\omega_1',\by_1') \overline{\hat{s}(\omega_2',\by_2')} \rangle - \langle \overline{\hat{s}(\omega_1,\by_1)} \hat{s}(\omega_2,\by_2) \rangle \langle\hat{s}(\omega_1',\by_1') \overline{\hat{s}(\omega_2',\by_2')} \rangle\\
&= (2\pi)^2 
\hat{F}(\omega_1)\hat{F}(\omega_2) K(\by_1) K(\by_2) \big[ 
 \delta(\omega_1'-\omega_1)\delta(\omega_2'-\omega_2) \delta( \by_1'-\by_1 ) \delta( \by_2'-\by_2)
 \\
 & \quad +
 \delta(\omega_2'+\omega_1)\delta(\omega_1'+\omega_2) \delta( \by_2'-\by_1 ) \delta( \by_1'-\by_2)
 \big] ,
  \end{align*}
we get the desired result (\ref{eq:varST1}).
\qed

A special case (which is not of direct interest for us) is when $\phi(t)=\psi(t)=\pi^{-1/2}\exp(-t^2)$ and $T'=2T$. Then 
$$
S_T(\omega) = \overline{\hat{u}_T(\omega,\bx_r)} \hat{u}_T(\omega,\bx_{r'}) ,
\quad \quad u_T(t,\bx) = \frac{1}{\sqrt{2\pi}} u(t,\bx) \exp \Big( - \frac{t^2}{2T^2}\Big) .
$$
The ECSD $S_T(\omega)$ is here the product of two correlated complex circular Gaussian variables
and it is not statistically stable (its standard deviation is of the same order as its expectation). 
In particular, for $\bx_{r'}=\bx_r$, the ECSD $S_T(\omega)$ follows an exponential distribution
because it is the square modulus of a  complex circular Gaussian variable. This indicates that it is necessary to take $T\gg T'$ to get 
a statistically stable quantity.\\

If $T'$ is fixed and of the order of $1/B$ and $T$ is much larger than $1/B$, 
 where $B$ is the bandwidth of the noise sources, then we get from (\ref{eq:varST1}):
\begin{align}
\nonumber
{\rm Var}(S_T(\omega)) =& 
\frac{\|\phi\|_{L^2}^2}{2\pi T} 
\int_\RR
\Big[ 
\hat{Q}(\omega_1,\bx_r,\bx_r) \hat{Q}(\omega_1,\bx_{r'},\bx_{r'}) 
\big|\hat{\psi} \big(   {T'} ( \omega- \omega_1 )\big) \big|^2
 \\
&+ |\hat{Q}(\omega_1,\bx_r,\bx_{r'})|^2
\hat{\psi} \big(   {T'} ( \omega- \omega_1) \big) 
\overline{\hat{\psi}}\big(   {T'} ( \omega + \omega_1) \big)
\Big] 
\hat{F}(\omega_1)^2
 d\omega_1 ,
\end{align}
where $\|\phi\|_{L^2}^2 = \int_\RR \phi(s)^2 ds$.
This shows that, if $T'$ is of the order of $1/B$, then the relative variance of the ECSD $S_T(\omega)$
is of the order of $1/(BT)$ and the ECSD is a statistically stable quantity.
We will see a quantitive analysis in Section \ref{sec:passivecomm}.

\section{Green's functions with resonant inclusions}
\label{sec:green}
The form of the Green's function $\hat{G}$ plays a fundamental role 
in the characterization of the cross covariance function of the 
wave field and the ECSD.
In the presence of non-dispersive inclusions, it has been known for a long time that it is possible to write small-volume expansions of the Green's function using  layer potential techniques \cite{ammarigarnier}.
The case of dispersive or resonant inclusions has been addressed more recently \cite{ammari18} and we 
briefly review in this section two different situations (in acoustics and electromagnetism) where the 
Green's function of a medium with resonant inclusions can be expanded in the general form (\ref{eq:green1}).
This gives an interesting form of the kernel $\hat{Q}$ as discussed in Subsection \ref{subsec:cc}.

We denote by $\hat{G}_0(\omega,\bx,\by)$ the homogeneous Green's function,
that is the  fundamental solution of the Helmholtz equation
\begin{equation}
\Delta_\bx \hat{G}_0 + \frac{\omega^2}{c_o^2}  \hat{G}_0 = - \delta(\bx-\by),
\end{equation}
subjected to Sommerfeld radiation conditions.
It is equal to 
\begin{align}
\hat{G}_0(\omega,\bx,\by)=\frac{1}{4\pi |\bx-\by|} \exp\Big( i \frac{\omega}{c_o} |\bx-\by|\Big).
\end{align}

\subsection{Acoustic waves in bubbly fluids}
We consider the acoustic scattering problem of a time-harmonic wave with frequency $\omega$
incident in water on a collection of $J$ air bubbles localized at $\bz_j$, $j=1,\ldots,J$. 
For simplicity, we assume that the bubbles are identical and spherical with radius $R$.
We denote by $\rho_o$ and $\kappa_o$, resp. $\rho_1$ and $\kappa_1$,  the density and bulk modulus of water, resp. air.
The sound speed in water, resp. air, is $c_o=\sqrt{\kappa_o / \rho_o}$, resp. $c_1=\sqrt{\kappa_1 / \rho_1}$. 
We also introduce the dimensionless parameters $\delta = \rho_1/\rho_o$, which is much smaller than one,  and 
$\alpha_1 = \omega R/c_1$.
If $\alpha_1\ll 1$,  if the minimal distance between the bubbles is larger than 
$|\rho(\omega)|$ (see (\ref{def:rhoomega})), and if the frequency $\omega$ is close to the Minnaert frequency (see (\ref{def:minnaert})), then the Green's function can be expanded as \cite{ammari17,ammari17b}
\begin{align}
\label{eq:green1a}
\hat{G}(\omega,\bx,\by) = &\hat{G}_0(\omega,\bx,\by)
+  \sum_{j=1}^J \rho(\omega)  \hat{G}_0(\omega,\bx,\bz_j)\hat{G}_0(\omega,\by,\bz_j) ,
\end{align}
where
\begin{align}
\label{def:rhoomega}
\rho(\omega)= 4 \pi R \frac{1-\alpha_1{\rm cot}(\alpha_1)}{\alpha_1{\rm cot}(\alpha_1)-1+\delta - i \delta (c_1/c_o) \alpha_1} .
\end{align}
The Minnaert frequency is 
\begin{equation}
\label{def:minnaert}
\omega_M= \frac{ c_1 \alpha_M}{R},
\end{equation}
where $\alpha_M$ is the solution of $\tan(\alpha_M) = \alpha_M/(1-\delta)$,
that is to say, $\alpha_M\simeq \sqrt{3\delta}$.
In particular, if $\omega=\omega_M$, then $\rho(\omega_M) = 4 \pi R  \frac{c_o}{c_1 \alpha_M} i$ is purely imaginary
and it is approximately equal to $880 R i$ for air bubbles in water ($\rho_o =10^6$~g.m${}^{-3}$, 
$\rho_1=1.29 \, 10^3$~g.m${}^{-3}$, $c_o=1482$~m.s${}^{-1}$, $c_1=340$~m.s${}^{-1}$ \cite{ammari18}).

\subsection{Electromagnetic waves with plasmonic nanoparticles}
We consider the electromagnetic scattering problem of a time-harmonic wave with frequency $\omega$ incident on a collection of $J$ plasmonic nanoparticles localized at $\bz_j$, $j=1,\ldots,J$. 
For simplicity, we use the Helmholtz equation instead of the full Maxwell equations and we assume that the nanoparticles are identical.
The background medium is homogeneous with electric permittivity $\eps_o$ and magnetic permeability $\mu_o$,
 while the particle occupying a bounded and simply connected domain ${\cal D}$ has  electric permittivity $\eps(\omega)$ and magnetic permeability $\mu(\omega)$, 
both of which may depend on the frequency.
A typical form for a nanoparticle is the Drude model \cite{sarid}:
\begin{align}
\eps(\omega)   = 
\eps_o \Big(1 - \frac{ \omega_p^2}{\omega^2 +  i\omega /\tau}\Big),
\quad\quad
\mu(\omega) = 
\mu_o \Big(1 - \frac{F_f \omega^2}{ \omega^2-\omega_r^2 + i \omega /\tau}\Big),
\end{align}
where $\tau > 0$ is the nanoparticle's bulk electron relaxation rate ($\tau^{-1}$ is the damping coefficient), $F_f$ is a filling factor, 
$\omega_p$ is the plasma frequency of the bulk material,
and $\omega_r$ is a localized plasmon resonant frequency. 
A small-volume expansion of the electromagnetic field using layer potential techniques
gives the following expansion of the Green's function 
\cite[Proposition 4.2]{ammari15}:
\begin{align}
\nonumber
\hat{G}(\omega,\bx,\by) = &\hat{G}_0(\omega,\bx,\by)
+  \sum_{j=1}^J \rho(\omega)  \hat{G}_0(\omega,\bx,\bz_j)\hat{G}_0(\omega,\by,\bz_j) \\
&
+ \sum_{j=1}^J
\nabla_\bz \hat{G}_0(\omega,\bx,\bz_j)^T {\bf M}(\omega) \nabla_\bz  \hat{G}_0(\omega,\by,\bz_j) ,
\label{eq:green1}
\end{align}
where
\begin{align}
\rho(\omega)  &= \frac{\omega^2}{c_o^2}  \Big(\frac{\eps(\omega)}{\eps_o} -1\Big) |{\cal D}|   ,
\end{align}
$|{\cal D}|$ is the volume of ${\cal D}$,
${\bf M}(\omega)$ is the polarization tensor, that is a symmetric complex matrix that depends on $\mu(\omega)$ and ${\cal D}$ and that is of order $|{\cal D}|$,
and the remainder is of order $|{\cal D}|^{4/3}$.

\subsection{Second-order moments of recorded signals}
\label{subsec:cc}

We consider
the kernel $\hat{Q}$ defined by (\ref{def:hatQ})  that determines the properties
of the cross covariance function of the wave field and the ECSD.

\begin{proposition}
\label{prop:expandgreen}
When the Green's function can be expanded in the form (\ref{eq:green1})
and the noise support function $K$ completely surrounds the region of interest containing the two receiver points $\bx_r$ and $\bx_{r'}$ and 
the inclusions at $\bz_j$,
then the kernel $\hat{Q}$ can be expanded in the form
\begin{align}
\nonumber
& \hat{Q}(\omega,\bx_r,\bx_{r'}) =
\frac{c_o}{\omega}{\rm Im}\big(\hat{G}_0(\omega,\bx_r,\bx_{r'}) \big)
\\
\nonumber
& \hspace*{0.3in}
+
\frac{c_o}{\omega}
 \sum_{j=1}^J  {\rm Im} \Big( \rho(\omega) \hat{G}_0(\omega,\bx_r,\bz_j) \hat{G}_0(\omega,\bx_{r'},\bz_j)  \Big) \\
\nonumber
& \hspace*{0.3in}
-
\frac{c_o}{\omega} \sum_{j=1}^J {\rm Im}\big(\rho(\omega)\big)  \overline{\hat{G}_0(\omega,\bx_r,\bz_j)}
\hat{G}_0(\omega,\bx_{r'},\bz_j)  \\
\nonumber
& \hspace*{0.3in}
+\frac{c_o}{\omega} \sum_{j=1}^J  {\rm Im} \Big( \nabla_\bz \hat{G}_0(\omega,\bx_r,\bz_j)^T {\bf M}(\omega) 
\nabla_\bz  \hat{G}_0(\omega,\bx_{r'},\bz_j) \Big)\\
& \hspace*{0.3in}
- \frac{c_o}{\omega} \sum_{j=1}^J   \nabla_\bz\overline{ \hat{G}_0(\omega,\bx_r,\bz_j)}^T {\rm Im} 
\big( {\bf M}(\omega) \big) \nabla_\bz  \hat{G}_0(\omega,\bx_{r'} ,\bz_j) .
\label{eq:expandQ}
\end{align}
\end{proposition}
Let us analyze the terms of the right-hand side of (\ref{eq:expandQ}).\\
\begin{itemize}
\item
The first term of the right-hand side comes from the standard Helmholtz-Kirchhoff identity (\ref{eq:standardhk}) applied to the homogeneous 
Green's function.
\item
The second and fourth terms come  from the standard Helmholtz-Kirchhoff identity (\ref{eq:standardhk}) applied to the correction of the Green's function
due to the inclusions:
\begin{align*}
&\frac{c_o}{\omega} 
\sum_{j=1}^J  {\rm Im} \Big( \rho(\omega) \hat{G}_0(\omega,\bx_r,\bz_j) \hat{G}_0(\omega,\bx_{r'},\bz_j)  \Big) \\
&+\frac{c_o}{\omega} \sum_{j=1}^J  {\rm Im} \Big( \nabla_\bz \hat{G}_0(\omega,\bx_r,\bz_j)^T {\bf M}(\omega) 
\nabla_\bz  \hat{G}_0(\omega,\bx_{r'},\bz_j) \Big) \\
&=\frac{c_o}{\omega}   {\rm Im} \Big(   \hat{G}(\omega,\bx_r,\bx_{r'} ) -  \hat{G}_0(\omega,\bx_r,\bx_{r'} ) \Big) ,
\end{align*}
by (\ref{eq:green1}).
These terms have been studied and used for applications to passive imaging \cite{agaltsov,noisebook,wap10}.
Indeed these terms show that (the imaginary part of) the full Green's function (\ref{eq:green1}) between the receiver points can be extracted from the 
cross covariance function of the noise signals recorded by them,
and that the data collected by a passive receiver array can be processed by cross correlation techniques to give the data 
that would be collected if the array were an active transducer array.
\item
The third and fifth terms of the right-hand side of Eq.~(\ref{eq:expandQ}) are usually not present in the literature \cite{ammarigarnier,noisebook,wap10}, because they can only appear when the medium is dispersive and
the coefficients of the Helmholtz equation are frequency-dependent with non-zero imaginary parts.
They can, however, be deduced from the general formulas that can be found in 
\cite{snieder07} and that extend the concept of the extraction of the Green's function to a wide class of scalar linear systems.
They can also be obtained from the generalized Helmholtz-Kirchhoff identity (\ref{eq:hk}) shown in Appendix \ref{sec:app}, after substitution of the expansion
(\ref{eq:green1}) of the Green's function $\hat{G}$.
These terms are very interesting from the theoretical point of view and from the practical point of view.
From the theoretical point of view, 
they are proportional to the imaginary parts of the reflectivity $\rho(\omega)$ and the polarization tensor ${\bf M}(\omega)$.
As a consequence, if the medium contains uncontrolled inclusions or reflectors that induce scattered waves
and one controlled dispersive device, then only the latter will appear in these terms of (\ref{eq:expandQ}).
Moreover, these terms involve a product of two Green's functions, one of them being complex-conjugated.
We know that such quantities are stable with respect to clutter noise (noise induced by a randomly scattering medium) \cite{book,noisebook},
and we will see interesting consequences of this property below.
From the practical point of view, these two terms are the motivation to propose an application to passive communication, in the same way as 
passive imaging has been developed in the last fifteen years from the inspection and the manipulation of the second and fourth terms of the right-hand side of (\ref{eq:expandQ}).
By exploiting these new terms we will show in the next section that it is possible to communicate between two passive arrays using only ambient noise illumination.
\end{itemize}

\noindent
{\it Proof.}
We address the case where the Green's function can be expanded as (\ref{eq:green1}) with ${\bf M}(\omega)={\bf 0}$
and the noise support function is uniformly supported at the surface of the ball $B({\bf 0},L)$ for a large $L$:
$$
K(\by) = \delta_{\partial B({\bf 0},L)} (\by).
$$
We then have
\begin{align*}
&\hat{Q}(\omega,\bx_r,\bx_{r'}) =
\int_{\partial B({\bf 0},L)} \overline{ \hat{G}_0 (\omega,\bx_r,\by) }\hat{G}_0 (\omega,\bx_{r'},\by) d\sigma(\by)
\\
&\quad +
\sum_{j=1}^J  \rho(\omega)
\int_{\partial B({\bf 0},L)} \overline{ \hat{G}_0 (\omega,\bx_r,\by) }\hat{G}_0 (\omega,\bx_{r'},\bz_j) \hat{G}_0 (\omega,\bz_j,\by) d\sigma(\by)
\\
&\quad +
\sum_{j=1}^J  \overline{\rho(\omega)}
\int_{\partial B({\bf 0},L)} \overline{ \hat{G}_0 (\omega,\bx_r,\bz_j) } \overline{\hat{G}_0 (\omega,\bz_j,\by)} 
 \hat{G}_0 (\omega,\bx_{r'},\by) d\sigma(\by) ,
\end{align*}
where we neglect the quadratic terms in $\rho$ consistently with the expansion (\ref{eq:green1}).
By applying the standard Helmholtz-Kirchhoff identity (\ref{eq:standardhk}), we get as $L\to +\infty$:
\begin{align*}
&\hat{Q}(\omega,\bx_r,\bx_{r'}) = \frac{c_o}{2i \omega}
\big( \hat{G}_0 (\omega,\bx_r,\bx_{r'}) - \overline{ \hat{G}_0 (\omega,\bx_r,\bx_{r'}) }\big)
\\
&\quad +
\sum_{j=1}^J  \frac{c_o  \rho(\omega)}{2i \omega}
\big( \hat{G}_0 (\omega,\bx_r,\bz_j) -\overline{ \hat{G}_0 (\omega,\bx_r,\bz_j) }\big)  \hat{G}_0 (\omega,\bx_{r'},\bz_j)
\\
&\quad +
\sum_{j=1}^J  \frac{c_o  \overline{\rho(\omega)}}{2i \omega}
\overline{ \hat{G}_0 (\omega,\bx_r,\bz_j) } 
\big(  \hat{G}_0 (\omega,\bx_{r'},\bz_j) - \overline{\hat{G}_0 (\omega,\bx_{r'},\bz_j) } \big),
\end{align*}
which gives the desired result.
The general case (\ref{eq:green1}) with ${\bf M}(\omega) \neq {\bf 0}$ can be addressed similarly.
\qed

\section{Passive communication}
\label{sec:passivecomm}

\subsection{Communication scheme}

We consider the following passive communication problem.
The problem is to transmit a binary message from a passive transmitter array (PTA) to a passive receiver array (PRA).
The medium is illuminated by ambient noise sources with narrowband spectrum concentrated around the frequency $\omega_o$.
We may think at the 4G  broadband cellular network technology, whose carrier frequency is around $\omega_o=2\pi \times 3\, 10^9$ rad.s${}^{-1}$
and the bandwidth is around $B=2\pi \times 10^7$   rad.s${}^{-1}$.

The PTA consists of $J$ small resonant devices localized at $\bz_j$, $j=1,\ldots,J$.
These devices can be tuned electronically so that the PTA can impose a desired value of the imaginary part of its 
electric permittivity, hence a desired value of the imaginary part of $\rho(\omega)$ for $\omega$ close to $\omega_o$.
Typically the PTA can impose ${\rm Im}(\rho(\omega_o)) \in \{0,\rho_1\}$ and we assume that ${\rm Im}(\rho(\omega))$ is close to ${\rm Im}(\rho(\omega_o))$ as long as $\omega -\omega_o$ is of the order of the bandwidth $B$.
The PTA can then transmit a binary message  $(\delta_k)_{k=0}^{K-1} \in \{0,1\}^K$  by encoding it as a step-wise constant
modulation 
$$
{\rm Im}(\rho(\omega_o))(t) = \sum_{k=0}^{K-1} \delta_k \rho_1 {\bf 1}_{ [(4k+2)T,(4k+4)T]} (t).
$$
This means that 
during the time intervals of the form $[4kT,(4k+2)T]$ (with duration $2T$), the PTA does not activate its elements so that ${\rm Im}(\rho(\omega_o))=0$.
During the time intervals of the form $[(4k+2)T,(4k+4)T]$, the PTA activates all its elements to impose ${\rm Im}(\rho(\omega_o))= \delta_k \rho_1$.

The PRA consists of two receivers at $\bx_r$ and $\bx_{r'}$. It computes the ECSD 
at carrier frequency $\omega_o$ and at successive times $(2k+1)T$ (separated by $2T$):
$$
{\cal S}_k =
   \iint_{\RR^2} u\big( t -\frac{\tau}{2}, \bx_r\big)u\big( t +\frac{\tau}{2}, \bx_{r'}\big)
e^{ i\omega_o \tau} \phi_T\big(t - (2k+1)T \big) \psi_{T'}(\tau) dt d\tau  , 
$$
for $k=0,\ldots,2K-1$.
Assuming that $T$ is large enough to ensure statistical stability,
we can anticipate from Propositions \ref{prop:momECSD} and \ref{prop:expandgreen} that ${\cal S}_k$ is the sum of three terms
corresponding to the terms in the expansion (\ref{eq:expandQ}) of $\hat{Q}$: \\
1) The first term
$$
\left< {\cal S}_k\right>_I 
=\frac{1}{2\pi} 
\int_\RR
\frac{c_o}{\omega_1}{\rm Im}\big(\hat{G}_0(\omega_1,\bx_r,\bx_{r'}) \big)
 \hat{F}(\omega_1) 
\hat{\psi}\big( T'(\omega_o-\omega_1)\big)
d\omega_1 
$$
 is constant in $k$ because it does not depend on $\rho$.\\
2) The second term is
\begin{align*}
\left< {\cal S}_k\right>_{II} =
\frac{1}{2\pi} 
\int_\RR
\frac{c_o}{\omega_1}
 {\rm Im} \Big( \rho(\omega_1)  \Big[ \sum_{j=1}^J  \hat{G}_0(\omega_1,\bx_r,\bz_j) \hat{G}_0(\omega_1,\bx_{r'},\bz_j)  \Big] \Big)
 \\
 \times
 \hat{F}(\omega_1) 
\hat{\psi}\big( T'(\omega_o-\omega_1)\big)
d\omega_1   ,
\end{align*}
where the value of $\rho(\omega_1)$ is taken during the time interval $[2kT,(2k+2)T]$.
The second term is negligible for large $J$ because the sum over $J$ averages out to zero (this is the sum of the product of two Green's functions which have large and different phases).\\
3) The third term 
\begin{align*}
\left< {\cal S}_k\right>_{III} 
=- \frac{1}{2\pi} 
\int_\RR
\frac{c_o}{\omega_1}
 {\rm Im} \big( \rho(\omega_1) \big)
 \Big[  \sum_{j=1}^J  \overline{\hat{G}_0(\omega_1,\bx_r,\bz_j)} \hat{G}_0(\omega_1,\bx_{r'},\bz_j)  \Big] \\
 \times
 \hat{F}(\omega_1) 
\hat{\psi}\big( T'(\omega_o-\omega_1)\big)
d\omega_1 
\end{align*}
is proportional to the imaginary part of $\rho$ during the time interval $[2kT,(2k+2)T]$
and the sum over $J$ has a non-negligible value (this is the sum of the product of two Green's functions, one of them being complex-conjugated,
so that their phases essentially cancel out).

The imaginary part of $\rho$  is $0$ for even $k$ and $\delta_{(k-1)/2} \rho_1$ for odd $k$.
As a result, it is possible to decode the binary message by computing
\begin{equation}
\label{def:Deltak}
\Delta_k ={\cal S}_{2k+1}-{\cal S}_{2k} ,
\end{equation}
for $k=0,\ldots,K-1$,
which removes the constant term and gives a quantity that is proportional to 
$\delta_k \rho_1$.

We can choose $T'=1/B$ to exploit the full noise bandwidth and $\phi(t)=(1-|t|)_+$, 
so that there is no cross-talk between the windows with duration $2T$.
Under these conditions the transmission rate is $1/(4T)$.
As we will see below $T$ has to be large enough to ensure statistical stability (i.e. $ {\cal S}_k \simeq \left< {\cal S}_k\right>$), which limits the transmission rate.
Moreover,
 $J$ has to be large enough to ensure that the third term $\left< {\cal S}_k\right>_{III}$  is larger than the second term $\left< {\cal S}_k\right>_{II}$.

To summarize, this scheme works well if the ECSD is indeed a constant term plus a term proportional to ${\rm Im}(\rho(\omega_o))$. This is what we analyze in detail in the next subsection.

\subsection{Analysis of the ECSD}
We assume that the noise sources are narrowband, in the sense that the spectrum is of the form (for $\omega>0$):
\begin{equation}
\hat{F}(\omega) =  \frac{1}{B} \hat{F}_0 \Big( \frac{\omega-\omega_o}{B}\Big) ,
\label{eq:formFF0}
\end{equation}
with $B \ll \omega_o$ and $\hat{F}_0$ a compactly supported, even, nonnegative real-valued function.

The ECSD is dominated by the contribution of ${\rm Im}(\hat{G}_0(\omega,\bx_r,\bx_{r'}))$ in $\hat{Q}(\omega,\bx_r,\bx_{r'})$.
That is why we use the special encoding described in the previous section which makes it possible to remove this contribution by substraction of two successive measured ECSD.
We remark, however, that it is possible to essentially mitigate the term ${\rm Im}(\hat{G}_0(\omega,\bx_r,\bx_{r'}))$ 
by placing the two receivers of the PRA at $\lambda_o/2$ apart from each other, because we then have 
$$
{\rm Im}(\hat{G}_0(\omega_o,\bx_r,\bx_{r'})) = 
\frac{ \sin ( \frac{\omega_o}{c_o} \frac{\lambda_o}{2})}{4\pi  \frac{\lambda_o}{2}} = 0, 
\mbox{ since }
\lambda_o=\frac{2\pi c_o}{\omega_o},
$$
and therefore $
{\rm Im}(\hat{G}_0(\omega,\bx_r,\bx_{r'})) $ is close to zero for $\omega$ in the source spectrum that is narrowband around $\omega_o$.
This choice is very advantageous from the signal-to-noise point of view
and it can be easily implemented in practice (in the cell phone application, the wavelength is typically $10$~cm).
This is the configuration that we want to study in detail.

\begin{figure}
\begin{center}
\vglue-0.8cm
\includegraphics[width=9.5cm]{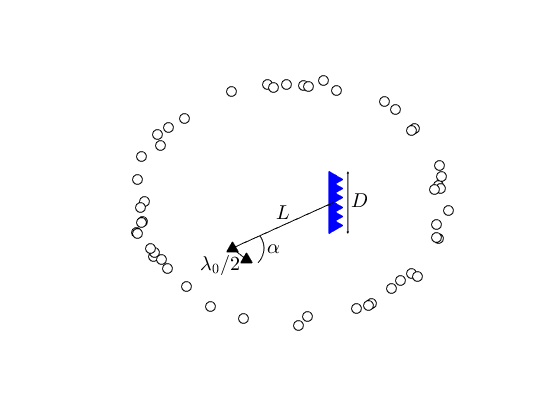}
\end{center}
\vglue-1.2cm
\caption{Geometric setup of the passive communication scheme.
The circles are the distant opportunistic sources.
The PTA is a planar array with diameter $D$.
The PRA has two receivers $\lambda_o/2$ apart from each other and at distance $L$ from the PTA.
}
\label{fig:setup}
\end{figure}

\begin{proposition}
\label{prop:meanST}
Let us consider the case where $T'=1/B$ and $|\bx_r-\bx_{r'}|=\lambda_o/2$,
where $\lambda_o=2\pi c_o/\omega_o$.
We assume that the Green's function can be expanded as (\ref{eq:green1}) with ${\bf M}={\bf 0}$ (for simplicity)
and we also assume that the inclusions at $\bz_j$ are in a planar array with diameter $D$ whose center $\bz_0$ is at the distance $L$ from $\bx_r$,
  with $\lambda_o \ll D\ll L$ (this is the paraxial approximation). We denote by $\alpha$ the angle between $\bx_{r'}-\bx_r$ and $\bz_0-\bx_r$ (see Figure \ref{fig:setup}).
We have
\begin{align}
\nonumber
\left< S_{T}(\omega_o) \right> =& 
\frac{1}{2^3 \pi^2}
 \Big[ \int_\RR \hat{F}_0(s) s^2 
 \hat{\psi}(s)
 ds \Big] \frac{B^2}{\omega_o^2}\\
&+ \frac{J \lambda_o}{2^6 \pi^4 L^2}  
\Big[ {\rm Im} \big(  R_{B1} R_{B2} e^{-i \pi \cos( \alpha)}  \big) 
-
 {\rm Im} (\rho_B) e^{-i \pi \cos( \alpha)}  \Big] ,
\end{align}
where
\begin{align}
\rho_B =&  \int_\RR\rho(\omega_o+Bs ) \hat{F}_0(s) 
 \hat{\psi}(s)
ds ,\\
\label{def:RB1}
R_{B1}=& \frac{1}{J} \sum_{j=1}^J 
\exp\big(2 i \frac{\omega_o}{c_o} |\bx_r-\bz_j| \big) , \\
R_{B2}=&
\int_\RR \rho(\omega_o+Bs) \exp\big(i \frac{2BL s}{c_o} \big)\hat{F}_0(s) 
 \hat{\psi}(s)
ds.
\end{align}
\end{proposition}

The term $|{\rm Im} \big(  R_{B1} R_{B2} e^{-i \pi \cos( \alpha)}  \big)|$ is much smaller than $| {\rm Im} (\rho_B)  |$
when $J$ is large.
Indeed, 
on the one hand, if $BL/c_o\ll 1$, then $R_{B2}\simeq \rho_B$ and if $BL/c_o\gg 1$, then $|R_{B2}|$ is small by Riemann-Lebesgue's lemma.
On the other hand, if $J$ is large, then 
the term $R_{B1}$ is small because it is the average of $J$ terms with modulus one and large and different phases (the phases are large because $\omega_o L /c_o \gg 1$), see 
Appendix \ref{app:B} and Lemma \ref{lem:tech} 
for a quantitative estimate.
As a result  $|R_{B1} R_{B2}|$ is much smaller than $| {\rm Im} (\rho_B)  |$.\\

\noindent
{\it Proof.}
We consider (\ref{eq:meanST1}) and use the form (\ref{eq:formFF0}) of the spectrum:
\begin{align*}
\left< S_T(\omega_o)\right> 
= 
\frac{1}{2\pi} \int_\RR \hat{Q}(\omega_o+Bs,\bx_r,\bx_{r'}) \hat{F}_0(s) 
\hat{\psi}(s) ds.
\end{align*}
By Proposition \ref{prop:expandgreen} (with ${\bf M}={\bf 0}$) 
it is the sum of three terms.
The first term is
$$
\left< S_T(\omega_o)\right>_I
= 
\frac{1}{2\pi} \int_\RR \frac{c_o}{\omega_o+Bs} {\rm Im}  \big(\hat{G}_0 (\omega_o+Bs,\bx_r,\bx_{r'}) \big)\hat{F}_0(s) 
\hat{\psi}(s) ds .
$$
Using the fact that $|\bx_r-\bx_{r'}| = \lambda_o/2$,
  we find
\begin{align*}
\left< S_T(\omega_o)\right> _I
= -
\frac{1}{8\pi^3} 
\int_\RR \frac{\sin( \pi \frac{B}{\omega_o} s)}{1+\frac{B}{\omega_o} s}
 \hat{F}_0(s)
 \hat{\psi}(s)
  ds  .
\end{align*}
As $B\ll \omega_o$ and $s\mapsto  \hat{F}_0(s)
 \hat{\psi}(s)$ is even, we get
\begin{align*}
\left< S_T(\omega_o)\right> _I
= 
 \frac{B^2}{8 \pi^2 \omega_o^2} 
 \int_\RR s^2
 \hat{F}_0(s)  \hat{\psi}(s)
 ds  .
\end{align*}
The second term is
\begin{align*}
\left< S_T(\omega_o)\right>_{II}
= 
\frac{1}{2\pi} \int_\RR \frac{c_o}{\omega_o+Bs} {\rm Im}  \Big(
 \sum_{j=1}^J  \rho(\omega_o+Bs) \hat{G}_0(\omega_o+Bs,\bx_r,\bz_j)\\
 \times \hat{G}_0(\omega_o+Bs,\bx_{r'},\bz_j)  \Big)
  \hat{F}_0(s) \hat{\psi}(s)
 ds  .
\end{align*}
Using $B\ll \omega_o$ and $|\bx_r-\bz_j|+|\bx_{r'}-\bz_j| =
2|\bx_r-\bz_j| - \lambda_o \cos (\alpha)/2+o(\lambda_o)$ because $\cos(\alpha) = (\bx_{r'}-\bx_{r}) \cdot (\bz_0-\bx_r) / [|\bx_r-\bx_{r'}| |\bz_0-\bx_r|]$,
we get
\begin{align*}
&\left< S_T(\omega_o)\right>_{II}
=
\frac{\lambda_o}{2^6 \pi^4 L^2} 
{\rm Im}  \big( R_B  \big) ,\\
&R_B =
\sum_{j=1}^J  \int_\RR 
\frac{\omega_o}{\omega_o+Bs}
\rho(\omega_o+Bs) \exp\Big(i \frac{\omega_o+Bs}{c_o} (|\bx_r-\bz_j|+|\bx_{r'}-\bz_j|)\Big) \hat{F}_0(s)
 \hat{\psi}(s) ds \\
&\hspace*{0.22in}  \simeq J R_{B1}R_{B2} \exp\big( - i\pi\cos(\alpha)\big)  .
\end{align*}
The third term is
\begin{align*}
\left< S_T(\omega_o)\right>_{III}
= -
\frac{1}{2\pi} \int_\RR \frac{c_o}{\omega_o+Bs}
 \sum_{j=1}^J {\rm Im}\big(  \rho(\omega_o+Bs) \big)  \overline{\hat{G}_0(\omega_o+Bs,\bx_r,\bz_j)}\\
 \times \hat{G}_0(\omega_o+Bs,\bx_{r'},\bz_j) 
  \hat{F}_0(s) \hat{\psi}(s)
 ds .
\end{align*}
Using $B\ll \omega_o$ and $-|\bx_r-\bz_j|+|\bx_{r'}-\bz_j| = - \lambda_o \cos (\alpha)/2+o(\lambda_o)$,
we get
\begin{align*}
& \left< S_T(\omega_o)\right>_{III}
= -
\frac{\lambda_o}{2^6 \pi^4 L^2}  \tilde{R}_B ,\\
& \tilde{R}_B =
\sum_{j=1}^J  \int_\RR \frac{\omega_o}{\omega_o+Bs} {\rm Im}\big(\rho(\omega_o+Bs)\big) \exp\Big(i \frac{\omega_o+Bs}{c_o} (-|\bx_r-\bz_j|+|\bx_{r'}-\bz_j|)\Big) \\
&\hspace*{3.8in} \times  \hat{F}_0(s)
 \hat{\psi}(s) ds \\
&\hspace*{0.22in}  \simeq J {\rm Im}\big( \rho_B \big) \exp\big(- i \pi \cos( \alpha)\big) .
\end{align*}
This gives the desired result.
\qed

\begin{proposition}
\label{prop:varST}
Under the same conditions as in Proposition \ref{prop:meanST},
if $BT \gg 1$, then we have 
\begin{align}
\label{eq:varST2}
{\rm Var}\big( S_{T}(\omega_o) \big) = 
\frac{\|\phi\|_{L^2}^2}{2^5 \pi^3 BT}  \Big[\int_\RR \hat{F}_0(s)^2 |  \hat{\psi}(s)|^2 ds\Big]  
.
\end{align}
\end{proposition}

\noindent
{\it Proof.}
We consider (\ref{eq:varST1}) and use the form (\ref{eq:formFF0}) of the spectrum. 
As ${\rm Im}\big( \hat{G}_0(\omega,\bx_r,\bx_{r}) \big) = 
 {\omega}/({4\pi c_o})$, the leading-order term is
\begin{align*}
\nonumber
{\rm Var}\big( S_{T}(\omega_o) \big) = 
 \frac{1}{2^6\pi^4} 
 \iint_{\RR^2}
  \big| \hat{\psi}( \frac{s_1+s_2}{2}\big)\big|^2 \big| \hat{\phi}\big( BT(s_1-s_2)\big)\big|^2
   \hat{F}_0(s_1)\hat{F}_0(s_2) 
 ds_1 ds_2  ,
 \end{align*}
 which is equal to (\ref{eq:varST2}) for $BT \gg 1$.
\qed

Propositions \ref{prop:meanST} and \ref{prop:varST}
show that, if
\begin{equation}
\label{eq:cond}
\frac{1}{\sqrt{BT}} \ll  \frac{J \lambda_o}{L^2} \rho_1 ,
\end{equation}
 then the PRA can extract the binary information because $\Delta_k$ defined by (\ref{def:Deltak}) is then approximately equal to 
$$
\Delta_k \simeq -
\delta_k   \frac{J \lambda_o}{2^6 \pi^4 L^2}   {\rm Im} (\rho_B) e^{-i \pi \cos( \alpha)}  +O\Big(  \frac{1}{\sqrt{BT}}\Big) .
$$
The condition (\ref{eq:cond}) ensures that the signal-to-noise ratio (SNR) is good enough.
It shows that the SNR can be improved by adding elements to the PRA (i.e., by increasing $J$),
or by increasing $T$. By choosing a value of the SNR larger than one (say, $\sqrt{10}$), we can choose the value of $T$ which determines the transmission rate $1/(4T)$.
Typically, $\rho_B \sim \lambda_o$, so we can anticipate a transmission rate of the order of 
$B J^2 \lambda_o^4 /(10L^4)$.
In the cell phone application, $\lambda_o\sim 10$~cm, $L \sim 10$~m, $B / (2\pi)\sim 10$~MHz, 
and the smart reflecting surfaces have nowadays typically $J \sim 100$ elements \cite{zhao20}, so the transmission rate is $\sim 1$kbit/s.
This is a rather low transmission rate, as in backscatter communication systems \cite{liu13,xu18}, but this is sufficient for most IoT applications.

{\bf Technical remark.}
It may happen that the ECSD is measured by taking the  difference of the Power Spectral Densities of $u(t,\bx_r)+u(t,\bx_{r'})$ and 
$u(t,\bx_r)-u(t,\bx_{r'})$.
We then obtain ${\rm Re}({\cal S}_T(\omega_o))$
instead of  ${\cal S}_T(\omega_o)$.
Under such circumstances, it is necessary that $\cos (\pi \cos(\alpha))\neq 0$, i.e. $|\alpha| \not\in \{\pi/3,2\pi/3\}$, so that the third term in ${\rm Re}({\cal S}_T(\omega_o))$
is still dominant.
The condition $|\alpha| \not\in \{\pi/3,2\pi/3\}$ means that the two receivers $\bx_r$ and $\bx_{r'}$ should not make an angle of $\pi/3$ or $2\pi/3$ with respect to the PTA.
In order to ensure that the condition $|\alpha| \not\in \{\pi/3,2\pi/3\}$ is always fulfilled for at least a pair of receivers, it would be necessary to build a PRA with  four receivers 
disposed at the vertices of a tetrahedron.
Adding receivers is also a way to increase the SNR and therefore to increase the transmission rate,
but one usually wants cheap devices for  IoT applications.

\subsection{Stability with respect to measurement noise}
We study here the impact of additive measurement noise.
We consider that the recorded data is of the form
\begin{align}
u_{\rm meas} (t,\bx_r) = u(t,\bx_r) + \eps_{{\rm meas},r}(t) ,
\end{align}
where the additive noises $\eps_{{\rm meas},r}(t)$ are independent and identically distributed in $r$, stationary in time with mean zero and covariance function $C_{\rm meas}(t)$.
The measured ECSD $S_{{\rm meas},T}(\omega)$ defined as (\ref{def:ST}) with  $u_{\rm meas}$ in place of $u$
has mean 
\begin{align}
\left< S_{{\rm meas},T}(\omega)\right> =
\left< S_{T}(\omega)\right>
\end{align}
and variance
\begin{align}
\nonumber
{\rm Var}(S_{{\rm meas},T}(\omega)) 
=&
{\rm Var}(S_{T}(\omega)) 
+
 \iint_{\RR^4} {C}_{\rm meas}\big(t-t'+\frac{\tau-\tau'}{2} \big) {C}_{\rm meas}\big(t-t'-\frac{\tau-\tau'}{2} \big) 
\\
 \nonumber
&
 \times
e^{i \omega (\tau-\tau')} {\phi}_T(t)\phi_{T}(t') 
\psi_{T'}(\tau)\psi_{T'}(\tau') dt dt' d\tau d\tau' \\
\nonumber
=&
{\rm Var}(S_{T}(\omega)) 
+
\frac{1}{(2\pi)^2}
 \iint_{\RR^2} \hat{C}_{\rm meas}(\omega_1) \hat{C}_{\rm meas}(\omega_2) \\
&
 \times
\big|\hat{\phi}\big( T (\omega_1-\omega_2)\big)\big|^2  \big| \hat{\psi}\big(T'( \omega-\frac{\omega_1+\omega_2}{2})\big)\big|^2 
d\omega_1d\omega_2   .
\end{align}
If the additive noise has small coherence time $C_{\rm meas}(t)=\sigma_{\rm meas}^2 C_0(t/t_{\rm meas})$, with $C_0(0)=1$, $\int_\RR C_0(s) ds=1$, and $t_{\rm meas}\ll T'$, then we have
\begin{align}
{\rm Var}(S_{{\rm meas},T}(\omega)) 
=
{\rm Var}(S_{T}(\omega)) 
+
\|\phi\|_{L^2}^2\|\psi\|_{L^2}^2 \sigma_{\rm meas}^4  \frac{t_{\rm meas}^2}{T T'} .
\end{align}
If we take $T'=1/B$, then the additional variance is proportional to $  \frac{(Bt_{\rm meas})^2}{ BT}$.
Since $Bt_{\rm meas}\ll 1$, and we have already assumed that $BT\gg 1$, this shows that measurement noise is efficiently mitigated
by the communication scheme through the calculation of the ECSD.

\subsection{Stability with respect to clutter noise}
Here we consider the case where the medium contains a PTA (or localized inclusions with dispersive properties) and
a randomly heterogeneous background, with non-dispersive properties, with index of fraction $n^2 (\bx) = 1+ \nu(\bx)$ where $\nu$ is a zero-mean random process,
compactly supported in some (possibly large) domain.
The analysis carried out in this paper is still valid after replacement of the homogeneous Green's function $\hat{G}_0$ by the cluttered 
Green's function $\hat{G}_{\rm clu}$, that is, the Green's function of the randomly heterogeneous medium:
\begin{equation}
\Delta_\bx \hat{G}_{\rm clu}(\omega,\bx,\by) +\frac{\omega^2}{c_o^2}\big(1+\nu(\bx)\big) \hat{G}_{\rm clu}(\omega,\bx,\by)  =- \delta(\bx-\by).
\end{equation}
This is because the standard Helmholtz-Kirchhoff identity (\ref{eq:standardhk}) is also valid when the
medium is spatially varying as long as the coefficients of the Helmholtz equation are real-valued.
Under such circumstances, 
the kernel $\hat{Q}$ is given by (\ref{eq:expandQ}) in Proposition \ref{prop:expandgreen} with $\hat{G}_0$ replaced by $\hat{G}_{\rm clu}$.
The key observation is that the important term for passive communication 
\begin{equation}
\frac{c_o}{\omega} \sum_{j=1}^J {\rm Im}(\rho(\omega))  \overline{\hat{G}_{\rm clu}(\omega,\bx_r,\bz_j)}
\hat{G}_{\rm clu}(\omega,\bx_{r'},\bz_j)  
\end{equation}
involves a product of two cluttered Green's functions, one of them being complex conjugated, and $\bx_r$ and $\bx_{r'}$ are close to each other,
at half-a-wavelength apart from each other.
As a result the value of the random term $\overline{\hat{G}_{\rm clu}(\omega,\bx_r,\bz_j)}
\hat{G}_{\rm clu}(\omega,\bx_{r'},\bz_j)  $ is statistically stable (its realizations are close to its mean) and it is very close
to $\overline{\hat{G}_0(\omega,\bx_r,\bz_j)}
\hat{G}_0(\omega,\bx_{r'},\bz_j)  $
\cite{borcea11,book,garniersolna16,noisebook}.
The statistical stability property can be quantified by studying the expectation and the variance of 
$\overline{\hat{G}_{\rm clu}(\omega,\bx_r,\bz_j)}
\hat{G}_{\rm clu}(\omega,\bx_{r'},\bz_j)  $.
This study involves  the calculations of fourth-order moments of the cluttered Green's function \cite{borcea11,noisebook} and it shows that the random value of
$\overline{\hat{G}_{\rm clu}(\omega,\bx_r,\bz_j)}
\hat{G}_{\rm clu}(\omega,\bx_{r'},\bz_j)  $
is close to the deterministic value of $\overline{\hat{G}_0(\omega,\bx_r,\bz_j)}
\hat{G}_0(\omega,\bx_{r'},\bz_j)  $
 when the distance from $\bx_r$ to $\bx_{r'}$ is smaller than the scattering mean free path:
\begin{align*}
\EE\big[ \overline{\hat{G}_{\rm clu}(\omega,\bx_r,\bz_j)}
\hat{G}_{\rm clu}(\omega,\bx_{r'},\bz_j) \big]
& \simeq \overline{\hat{G}_0(\omega,\bx_r,\bz_j)}
\hat{G}_0(\omega,\bx_{r'},\bz_j) ,\\ 
{\rm Var} \big( \overline{\hat{G}_{\rm clu}(\omega,\bx_r,\bz_j)}
\hat{G}_{\rm clu}(\omega,\bx_{r'},\bz_j) \big) 
&\ll \big| \overline{\hat{G}_0(\omega,\bx_r,\bz_j)}
\hat{G}_0(\omega,\bx_{r'},\bz_j)\big|^2.
\end{align*}
The  scattering mean free path is usually larger than the wavelength, except in very strongly scattering media.
The statistical stability property in weakly scattering media  can be explained by the fact that the main effect of the cluttered medium is to add random phases $\phi_{\rm clu}(\omega, \bx_r,\bz_j)$ to the Green's function and that the random phases $\phi_{\rm clu}(\omega, \bx_r,\bz_j)$ and $\phi_{\rm clu}(\omega, \bx_{r'},\bz_j)$ are strongly correlated provided $\bx_r$ and $\bx_{r'}$ are close to each other, so that the random phases cancel each other in the product $\overline{\hat{G}_{\rm clu}(\omega,\bx_r,\bz_j)}
\hat{G}_{\rm clu}(\omega,\bx_{r'},\bz_j)  $.
This analysis shows that the passive communication scheme is expected to be very robust with respect to clutter noise.

\section{Conclusions and perspectives}

In this paper we have demonstrated that passive communication using narrowband ambient noise illumination and tunable metasurfaces is possible.
From the theoretical point of view the key arguments are a generalized version of the Helmholtz-Kirchhoff identity and
the statistical stability of the empirical cross spectral density of the wave field.
From the pratical point of view, the transmission rate that can be achieved with an illumination with carrier frequency of the order of a few GHz 
and a bandwidth of the order of $10$ MHz is of the order $1$ kbit/s.

We have addressed the case of full aperture ambient noise illumination, that is to say, we have assumed that the noise sources completely surround the region of interest.
This hypothesis is necessary to prove the efficiency of passive communication schemes by the generalized Helmholtz-Kirchhoff identity.
In other words, the proof of concept presented in this paper requires full aperture ambient noise illumination, but
we anticipate that it may be extended beyond this framework.
More exactly,
the results could certainly be extended to the case of partial aperture  ambient noise illumination in a reverberating cavity and in a strongly scattering medium, which can achieve wave equipartition \cite{bardos08,noisebook}.
The case of partial aperture  ambient noise illumination in open media or weakly scattering media deserves more study.
The overall principle would be similar but Proposition \ref{prop:expandgreen}
does not hold anymore and one should then use stationary phase arguments \cite{noisebook}
or semi-classical analysis \cite{bardos08} to study the structure of the kernel $\hat{Q}$.
Resolution properties such as the ones reported in \cite{reso}
would be necessary in order to determine the appropriate locations of the receiver antennas.
We can also anticipate that a randomly scattering medium can enhance the illumination diversity and make
the situation with partial aperture illumination closer to the full aperture case
when scattering is strong enough.

\appendix

\section{The generalized Helmholtz-Kirchhoff identity} \label{sec:app}
We consider a general form of the time-harmonic  three-dimensional scalar wave equation (or Helmholtz equation) in a dispersive and inhomogeneous medium:
\begin{equation}
\label{eq:helm}
\nabla_\bx \cdot \big( (1+ a(\omega,\bx)) \nabla_\bx \hat{u} \big) + \frac{\omega^2}{c_o^2} (1+b(\omega,\bx)) \hat{u} = - \hat{s}(\omega,\bx)   .
\end{equation}
The parameters $a$ and $b$ of the medium are frequency-dependent and complex-valued,
they are spatially varying and we assume that $a$ and $b$ are compactly supported in $\RR^3$.
This means that the medium is homogeneous with speed of propagation $c_o$ at infinity and 
the Helmholtz equation (\ref{eq:helm}) is well-posed with Sommerfeld radiation conditions:
\begin{equation}
\label{eq:sommerfeld}
\lim_{|\bx| \to \infty} |\bx| \Big( \frac{\bx}{|\bx|}  \cdot \nabla_\bx -
\ i\frac{\omega}{c_o} \Big)
\hat{u}( \omega,\bx ) = 0 .
\end{equation}

\begin{proposition}
We denote by $\hat{G}(\omega,\bx,\by)$ the Green's function of (\ref{eq:helm}), i.e. the unique solution of (\ref{eq:helm}) with $\hat{s}(\omega,\bx)=\delta(\bx-\by)$
satisfying the Sommerfeld radiation condition (\ref{eq:sommerfeld}).\\
{\it 1.}
The Green's function satisfies the reciprocity property $\hat{G}(\omega,\bx,\by) = \hat{G}(\omega,\by,\bx)$.\\
{\it 2.}
The Green's function satisfies the generalized Helmholtz-Kirchhoff identity: 
\begin{align}
\nonumber
& \lim_{L \to +\infty} \frac{\omega}{c_o}   \int_{\partial B({\bf 0},L)} 
\overline{ \hat{G}( \omega,\bx,\bz)}  \hat{G}( \omega ,\by,\bz)  d \sigma(\bz) 
=  {\rm Im} \big( \hat{G}(\omega, \bx,\by) \big)
\\
\nonumber
& \quad \quad- \frac{\omega^2}{c_o^2}
\int_{\RR^3} {\rm Im}(b(\omega,\bz) ) \overline{ \hat{G}( \omega,\bx,\bz)}  \hat{G}( \omega ,\by,\bz) d \bz\\
&
\quad \quad +
\int_{\RR^3} {\rm Im}(a(\omega,\bz) ) \nabla_\bz \overline{ \hat{G}( \omega,\bx,\bz)}\cdot\nabla_\bz  \hat{G}( \omega ,\by,\bz) d\bz
.
\label{eq:hk}
\end{align}
\end{proposition}
Of course, if $a$ and $b$ are real-valued, then
we recover the standard Helmholtz-Kirchhoff identity (\ref{eq:standardhk}).
This is the case when the medium is frequency-independent, or more generally when the medium is dispersive but invariant by time reversal,
so that $a$ and $b$ are even real-valued functions of $\omega$.\\

\noindent
{\it Proof.}
The proof of the reciprocity is exactly the same one as in the non-dispersive case \cite[Proposition 2.1]{noisebook}, it is based on the divergence theorem and the Sommerfeld radiation condition.
The proof of the generalized Helmholtz-Kirchhoff identity follows the lines of the proof of the standard Helmholtz-Kirchhoff identity \cite[Theorem 2.2]{noisebook} as we now explain.
Let us consider the Helmholtz equations satisfied by the Green's functions with point sources at $\bx$ and $\by$, 
with $\bx\neq \by$:
\begin{eqnarray*}
\nabla_\bz \cdot\big( (1+a(\omega,\bz)) \nabla_\bz
 \hat{G}(\omega,\bz,\by) \big)+ \frac{\omega^2}{c^2_o} (1+b(\omega, \bz))\hat{G}(\omega,\bz,\by) 
 =-\delta( \bz-\by),\\
\nabla_\bz \cdot\big( (1+\overline{a(\omega,\bz)}) \nabla_\bz
 \overline{\hat{G}(\omega,\bz,\bx)}\big)+ \frac{\omega^2}{c^2_o} (1+\overline{b(\omega, \bz)}) \overline{\hat{G}(\omega,\bz,\bx) }
 =-\delta( \bz-\bx).
\end{eqnarray*}
We multiply  the first equation  by $ \overline{\hat{G}(\omega,\bz,\bx)}$
and we subtract the second equation multiplied by  $\hat{G}(\omega,\bz,\by)$:
\begin{align*}
&\nabla_\bz \cdot \Big[  \overline{\hat{G}(\omega,\bz,\bx)}  \big(1+{a(\omega,\bz)}\big)
 \nabla_\bz  \hat{G}(\omega,\bz,\by)
-
 \hat{G}(\omega,\bz,\by)  \big(1+\overline{a(\omega,\bz)}\big)
 \nabla_\bz
  \overline{\hat{G}(\omega,\bz,\bx)} \big)
 \Big]\\
 &-\big( a(\omega,\bz)-\overline{a(\omega,\bz)}\big) \nabla_\bz \overline{\hat{G}(\omega,\bz,\bx)} \cdot \nabla_\bz \hat{G}(\omega,\bz,\by) \\
  &+\frac{\omega^2}{c_o^2} \big( b(\omega,\bz)-\overline{b(\omega,\bz)}\big) \overline{\hat{G}(\omega,\bz,\bx)} \hat{G}(\omega,\bz,\by) \\
 & = 
 \hat{G}(\omega,\bz,\by) \delta(\bz-\bx)
 -
  \overline{\hat{G}(\omega,\bz,\bx)} \delta(\bz-\by).
\end{align*}
Note the presence of the terms proportional to $a-\overline{a}$ and $b-\overline{b}$, that are absent in the non-dispersive case because $a$ and $b$ are then real-valued.
We integrate over the 
domain $B({\bf 0},L)$, for $L$ large enough so that the ball $B({\bf 0},L)$ contains $\bx$, $\by$, and the spatial supports of the functions $a$ and $b$.
By using the divergence theorem, we obtain
 \begin{align*}
&\int_{\partial B({\bf 0},L)} {\itbf n} (\bz)\cdot \Big[ \overline{\hat{G}(\omega,\bz,\bx)}  
 \nabla_\bz  \hat{G}(\omega,\bz,\by)
-
 \hat{G}(\omega,\bz,\by) 
 \nabla_\bz
  \overline{\hat{G}(\omega,\bz,\bx)} \big)
 \Big] d\sigma(\bz) \\
 &- \int_{ B({\bf 0},L)}\big( a(\omega,\bz)-\overline{a(\omega,\bz)}\big) \nabla_\bz \overline{\hat{G}(\omega,\bz,\bx)} \cdot \nabla_\bz \hat{G}(\omega,\bz,\by) d\bz \\
  &+\int_{ B({\bf 0},L)} \frac{\omega^2}{c_o^2} \big( b(\omega,\bz)-\overline{b(\omega,\bz)}\big) \overline{\hat{G}(\omega,\bz,\bx)} \hat{G}(\omega,\bz,\by) d \bz\\
 & = 
 \hat{G}(\omega,\bx,\by) 
 -
  \overline{\hat{G}(\omega,\by,\bx)} ,
\end{align*}
where ${\itbf n} (\bz) = \bz/|\bz|$ is the unit outward normal vector.
 As $a$ and $b$ are compactly supported in $B({\bf 0},L)$, 
the volume integrals over $B({\bf 0},L)$ in the second and third terms can be replaced by volume integrals over $\RR^3$.
By using the reciprocity property
and  the fact that 
the Green's function satisfies the Sommerfeld radiation condition
$$
\lim_{|\bz| \to \infty} |\bz| \Big( \frac{\bz}{|\bz|}  \cdot \nabla_\bz -
\ i\frac{\omega}{c_o} \Big)
\hat{G}( \omega,\bz, \bx ) = 0 ,
$$
we get  the desired result (\ref{eq:hk}) by letting $L\to+\infty$ in the surface integral.
\qed

\section{A technical result}
\label{app:B}
The goal of this section is to show that the term $R_{B1}$ defined by (\ref{def:RB1})
is small when $J$ is large. It is possible to address different configurations, where the PTA is dense or not, 
where the array geometry is regular or not, etc.
In the following lemma we address the case of a regular and dense PTA, which is the typical configuration of smart reflecting surfaces that are built nowadays~\cite{zhao20}.

\begin{lemma}
\label{lem:tech}
We consider the conditions of Proposition \ref{prop:meanST}.
We assume that the PTA is a dense regular array in the square plane with side length ${D}$.
We assume that the PRA is at distance $L$ from the PTA and that the angle between the normal direction to the PRA and the direction from the PTA to the PRA is small. 
We assume that  $L^2 \gg {D}^2 \gg L \lambda_o$.
We then have 
\begin{equation}
\label{eq1:lemB1}
|R_{B1}| \leq \frac{4 \lambda_oL}{ \pi {D}^2} .
\end{equation}
\end{lemma}

{\it Proof.}
We denote $L=|\bx_r - \bz_0|$ where $\bz_0$ is at the center of the PTA.
By expanding the difference $
 |\bx_r-\bz_j| - |\bx_r-\bz_0|$, we get
$$
2 |\bx_r-\bz_j|  =2 L-  2 \frac{ ( \bx_r -\bz_0) \cdot (\bz_j-\bz_0)  }{L}
+  \frac{ |\bz_j-\bz_0|^2  }{ L} + \cdots
 ,
 $$
so we find
\begin{align*}
& |R_{B1} |  \simeq  \Big| \frac{1}{J} \sum_{j=1}^J 
\exp\big(- 2 i \frac{\omega_o ( \bx_r -\bz_0) \cdot (\bz_j-\bz_0)  }{c_o L}  +  i \frac{\omega_o  |\bz_j-\bz_0|^2  }{c_o L} \big) \Big| \\
&=
\prod_{k=1}^2 \Big| \frac{1}{{D}}  \int_{-\frac{{D}}{2}}^{\frac{{D}}{2}} \exp\big( - 2i \frac{\omega_o x_k z}{c_o L} +i \frac{\omega_o z^2}{c_o L}  \big) dz \Big|
=
\prod_{k=1}^2 \Big| \frac{1}{{D}}  \int_{-\frac{{D}}{2}-x_k}^{\frac{{D}}{2}-x_k} \exp\big(i \frac{\omega_o z^2}{c_o L}  \big) dz \Big|\\
&=\frac{c_o L}{\omega_o {D}^2}  \prod_{k=1}^2 \Big|   C ( d_{k,+} )  - C(d_{k,-}) 
+i  S(d_{k,+})  - iS(d_{k,-}) \Big|  ,
\end{align*}
where $(x_1,x_2)$ are the coordinates of the projection of $\bx_r$ onto the plane of the PTA,
$C(d) = \int_0^d \cos(t^2)dt$ and $S(d)= \int_0^d \sin(t^2)dt$ are the Fresnel integrals \cite[Chapter 7]{abra}, and $d_{k,\pm} =\sqrt{ \frac{\omega_o}{c_o L}}\big(\pm \frac{{D}}{2} - x_k\big) $.
We get (\ref{eq1:lemB1}) by using the fact that $C$ and $S$ are bounded by one.
\qed

\end{document}